\documentclass[oneside,a4paper,11pt,shownumbers]{article}
\usepackage{latexsym}
\usepackage{euscript}
\usepackage{epsfig,amsmath,amssymb}

\def\be{\begin{equation}}
\def\ee{\end{equation}}
\def\bea{\begin{eqnarray}}
\def\eea{\end{eqnarray}}

\long\def\symbolfootnote[#1]#2{\begingroup%
\def\thefootnote{\fnsymbol{footnote}}\footnote[#1]{#2}\endgroup} 

 \large\normalsize

\begin{document}

\begin{center}

{\Large \bf Gravitating semilocal strings}

\vspace*{7mm} {Betti Hartmann $^{a}$
\symbolfootnote[1]{E-mail:b.hartmann@jacobs-university.de} and
Jon Urrestilla $^{b,c}$
\symbolfootnote[2]{E-mail:jon.urrestilla@ehu.es}}
\vspace*{.25cm}

${}^{a)}${\it School of Engineering and Science, Jacobs University Bremen, 28759 Bremen, Germany}\\
${}^{b)}${\it Department of Theoretical Physics, University of the Basque Country, UPV-EHU, 48080 Bilbao, Spain}\\
${}^{c)}${\it Department of Physics and Astronomy, University of Sussex, Brighton BN1 9QH, United Kingdom}\\

\vspace*{.3cm}
\end{center}

\begin{abstract}
We discuss the properties of semilocal strings minimally coupled to gravity. Semilocal strings are 
solutions of the bosonic sector of the Standard Model in the limit $\sin^2\theta_W=1$
(where $\theta_W$ is the Weinberg angle) and correspond to embedded Abelian--Higgs strings for a particular choice
of the scalar doublet.
We focus on the limit where the gauge boson mass is equal to the Higgs boson mass such that the solutions fulfill the Bogomolnyi-Prasad-Sommerfield (BPS) bound. 
\end{abstract}

\section{Introduction}
Semilocal strings \cite{av,hind} are solutions to the bosonic sector of the Standard Model
in the limit $\sin^2\theta_W=1$, where $\theta_W$ is the Weinberg angle. This model has $SU(2)\times U(1)$ symmetry, where the $SU(2)$ is global, while the $U(1)$ symmetry is a local, i.e. gauge symmetry. This model
can be seen as an extension of the $U(1)$ Abelian--Higgs model in which 
the complex scalar field of the Abelian--Higgs model is replaced by a complex scalar doublet, i.e. the number of
scalar degrees of freedom is doubled. 
Abelian--Higgs strings, however, have rather different properties than semilocal strings. As for the strings obtained in \cite{bdr}, the stability of semilocal strings does not follow from the topology of the vacuum manifold (as it does for
Abelian--Higgs strings), 
but from dynamical arguments. The ratio between the gauge and Higgs boson mass 
governs the stability of the semilocal strings: for Higgs boson mass larger (smaller) than the gauge boson mass semilocal strings are unstable (stable) and in  the case of equality between the masses (the BPS limit), a
degenerate one-parameter family of stable solutions exists. The parameter corresponds roughly to the width of the strings. In other words, in the BPS limit semilocal strings of arbitrary width
have the same energy. Therefore, there is a zero mode associated with the width of the semilocal strings. It was shown in \cite{leese} that when this zero mode gets excited, it always leads to the growth of the string core. 

It is interesting to note that semilocal strings share some properties with some new BPS strings. 
In \cite{bdr} the  conjecture that (stringy) D-strings are related to D-term BPS strings in 
the four dimensional field theory description was studied. New BPS cosmic strings of finite energy coupled to 
an axionic field were presented. The model has three different types of strings: 
tachyonic, axionic and hybrid strings. These objects are different from usual 
cosmic strings: among other things, a zero mode is present in the model, too.
This zero mode can also be excited  \cite{as} leading to decompactification effects.

Semilocal strings have been studied in cosmological settings both in the context of  their formation \cite{form}, network properties \cite{net} and their CMB implications \cite{UBHKL}. Due to an expected lower density of semilocal strings (as opposed to Abelian--Higgs strings) and due to the aforementioned tendency to become ``fatter'', these strings are interesting as possible solutions to relax constraints in inflationary models that predict too many cosmic strings \cite{UAD}.

In this paper we couple semilocal strings minimally to gravity. Besides obtaining the gravitating solutions, we will carefully study the properties of different members of the one-parameter families. As already mentioned, all solutions have the same (global) energy, but they have different energy densities. Therefore we expect to obtain differences in the local curvature of space--time. For example, in the supermassive case (when the symmetry breaking scale of the theory is larger than the Planck mass) Abelian--Higgs strings only exist up to some maximal value of the radial coordinate \cite{supermassive}. For the one-parameter families of semilocal strings, this maximal value
of the radial coordinate will change when varying the width of the string (and the gravitational coupling); and in the limit there will be a solution for all space.

\section{The model}
The action reads:
\begin{equation}
 S=\int d^4 x \ \sqrt{-g} \left( \frac{R}{16\pi G} + {\cal L}_M\right)  \ ,
\end{equation}
where $R$ denotes the Ricci scalar, $g$ is the determinant of the metric tensor, $G$ is Newton's constant and ${\cal L}_M$ denotes the Lagrangian density of the semilocal
model that has $SU(2)\times U(1)$ symmetry:
\begin{equation}
 {\cal L}_m=-\frac{1}{4} F_{\mu\nu} F^{\mu\nu} + (D_{\mu} \Phi)^{\dagger} D^{\mu} \Phi -\frac{\lambda}{2}\left(\Phi^{\dagger} \Phi - \eta^2\right)^2   \ .
\end{equation}
Here $F_{\mu\nu}=\partial_{\mu} A_{\mu} - \partial_{\nu} A_{\mu}$ is the field strength
tensor of the $U(1)$ gauge field and $D_{\mu} \Phi=(\partial_{\mu} - ie A_{\mu}) \Phi$ is the covariant
derivative of the complex scalar field doublet $\Phi=(\phi_1,\phi_2)^T$. $e$ denotes the gauge coupling,
$\lambda$ the self-coupling, $\eta$ the vacuum expectation value and $G$ Newton's constant. 
There are three mass scales in the theory: the Planck mass $M_{Pl}=G^{-1/2}$, the Higgs boson
mass $M_H=\sqrt{2\lambda} \eta$ and the gauge boson mass $M_W=\sqrt{2} e \eta$.
In the following, we will only be interested in the case with $M_H^2=M_W^2$, i.e. $e^2=\lambda$.
This is the Bogomolnyi--Prasad--Sommerfield (BPS) limit, that has been studied
extensively for $G=0$ in \cite{av,hind}.
In the following, we will keep $\lambda$ and $e^2$ explicitely, but keep in mind that
$e^2=\lambda$.

\subsection{Ansatz}
In this paper, we study solutions that are cylindrically symmetric, i.e. solutions that have a rotational
symmetry in the $x$-$y$-plane and do not explicitely depend on $z$.

The Ansatz for the metric in cylindrical coordinates ($t$, $\rho$, $\varphi$, $z$) is given by
\begin{equation}
 ds^2= N^2(\rho) dt^2 - d\rho^2 - L^2(\rho)d\varphi^2 - N^2(\rho)dz^2  \ ,
\end{equation}
where we have chosen $g_{tt}=-g_{zz}$ due to the boost symmetry of the solution.
Note that for BPS solutions, the Einstein equations tell us that $N(\rho)\equiv 1$
which is essentially related to the fact that the sum of the energy density $\epsilon=T_0^0$
and the pressure components $p_{\rho}$, $p_{\varphi}$, $p_z$ is zero in this case.
We will thus set $N(\rho)\equiv 1$ in the following - unless otherwise stated.

The Ansatz for the matter fields is \cite{av,hind}:
\begin{equation}
 A_{\varphi}= \frac{n}{e} a(\rho) \ \ , \ \ \phi_1= \eta f_1(\rho) e^{i n\varphi}  \ \ , \ \ 
\phi_2=\eta f_2(\rho) \ \ .
\end{equation}
By using the rescalings
$x=e\eta\rho$ and $L(x)=e\eta L(\rho)$
the energy density $\epsilon=T_0^0$ reads:
\begin{equation}
 \frac{T^0_0}{e^2\eta^4} = \frac{1}{2} \frac{1}{L^2} n^2 (a')^2 + (f_1')^2 + (f_2')^2 
+ \frac{1}{L^2} n^2 f_1^2 (1-a)^2 + \frac{1}{L^2} n^2 a^2 f_2^2 + \frac{\lambda}{2e^2} (f_1^2 + f_2^2 -1)^2  \ ,
\end{equation}
where now and in the following the prime denotes the derivative with respect to $x$.
The solutions fulfill an energy bound, such that the energy per unit length $\mu$ is given by~:
\begin{equation}
 \mu=\int d^2 x \sqrt{-g} T^0_0 = 2\pi \int dx L  T^0_0 = 2\pi n \eta^2  \ .
\end{equation}

The deficit angle is related to $\mu$ by
$\delta=8\pi G\mu = 8\pi G (2\pi n \eta^2)$
and is given by:
\begin{equation}
 \delta=2\pi (1-L'\vert_{\rho=\infty})  \ .
\end{equation}

\subsection{The equations}
The BPS equations read
\begin{eqnarray}
\label{fieldeq1}
 f_1' + \frac{n(a-1)}{L} f_1 =  0  \ \  ,  \ \ 
    f_2' + \frac{na}{L} f_2 =  0  \ \ , \ \
\frac{n a'}{L} + (f_1^2 +f_2^2 -1) =  0  
\end{eqnarray}
and the Einstein equation is
\begin{equation}
\label{fieldeq2}
 \frac{L''}{L}=-\alpha^2 \left((f_1^2+f_2^2-1)^2 + \frac{2n^2 (1-a)^2}{L^2}f_1^2 
+ \frac{2n^2 a^2}{L^2} f_2^2 \right)   \ ,
\end{equation}
where $\alpha^2 = (4\pi M_W^2)/(e^2 M_{Pl}^2)=  8\pi G\eta^2$.
Note that with this abbreviation, the deficit angle is given by
$\delta/(2\pi)=\alpha^2 n$.
For $\alpha=0$, the equations allow for a direct relation between the two scalar
field functions $f_1$, $f_2$ \cite{av,hind}. The relation for $\alpha\neq 0$ reads:
\begin{equation}
\label{relation}
 f_1=c\cdot \exp\left(\int \frac{n}{L} dx\right) \cdot f_2 \ \ .
\end{equation}
To solve the differential equations numerically, we have to impose appropriate boundary conditions.
The requirement of finiteness of the energy and regularity at the origin leads to the following conditions:
\begin{equation}
\label{bc}
 f_2'(0)=0 \ \ , \ \ a(0)=0 \ \ , \ \
L(0)=0 \ \ , \ \ L'(0)=1 \ \ , \ \ a(\infty)=1 \ \ .
\end{equation}
Note that we could also choose another set of boundary conditions for the matter fields by e.g.
imposing boundary conditions on $f_1$. These conditions are however equivalent to each other due to
the relation between the functions $f_1$ and $f_2$ (see below). The set of boundary conditions given in (\ref{bc})
has proven convenient in numerical simulations.
\section{Embedded Abelian--Higgs strings}
In the limit $f_2(\rho)\equiv 0$ the equations of motion (\ref{fieldeq1}), (\ref{fieldeq2})
reduce to the equations of the $U(1)$ Abelian--Higgs model minimally coupled to gravity.
Hence, the semilocal solutions correspond to 
embedded Abelian--Higgs strings. In contrast to the solutions
of the original $U(1)$ Abelian--Higgs model \cite{no}, the embedded Abelian--Higgs strings are however unstable in flat space--time for $M_H^2 > M_W^2$ \cite{hind}. 
In any case, the embedded strings share many properties with the usual $U(1)$ strings, it is therefore interesting to summarize the properties of gravitating
$U(1)$ Abelian--Higgs strings.
In \cite{linet1} it was shown that the solutions in curved space--time fulfill a BPS bound for $M_W=M_H$ that is essentially the same bound as that in flat space--time \cite{bogo}.
The explicit solutions have been constructed numerically in \cite{clv,bl}.
It has been observed that there are different types of solutions depending on the
choice of the gravitational coupling $\alpha$.

For $\alpha^2 n\leq 1$, the deficit angle is smaller than $2\pi$ and globally regular
solutions are possible. For $\alpha^2 n > 1$, the deficit angle is larger than $2\pi$
and the metric function $L$ vanishes at a finite value of the radial coordinate.
Since the deficit angle can only be larger than $2\pi$ if the symmetry breaking scale $\eta$ is larger than the Planck scale, these solutions were called ``supermassive'' strings \cite{supermassive}. 
 
Away from the BPS limit, the metric function $N(x)$ is non-trivial and it was realized
in \cite{clv,bl} that the string-like solutions have shadow solutions for the same
choice of Higgs to gauge boson mass ratio and gravitational
coupling.
For globally regular solutions, there are Melvin solutions in addition
to the string solutions which have a different asymptotic behaviour $N(x\rightarrow \infty)\propto x^{2/3}$ , $L(x\rightarrow\infty)\propto x^{-1/3}$, while for the supermassive solutions there are shadow solutions in the form of Kasner solutions.
For the latter, the metric function $N(x)$ vanishes at a finite value of the radial
coordinate, while $L(x)$ diverges there.
Note that Melvin and Kasner solutions do not exist in the BPS limit since $N(x)\equiv 1$.

\section{Semilocal strings}
In flat space-time, it was shown that solutions with $f_2(x)\neq 0$ exist only in the BPS limit $M_W=M_H$ \cite{av,hind}.
In fact, there exists a family of solutions that can be parametrized by the value of
$f_2(x)$ at the origin, $f_2(0)$, which can be interpreted as a condensate on the string.
For any choice of $f_2(0)$, the solutions fulfill the energy bound. They are hence degenerate
in energy and this degeneracy is directly linked to the existence of a zero mode.
Recently, semilocal strings coupled to so-called dark strings \cite{vachaspati} have been
studied \cite{bh2} and it was shown that semilocal solutions with $f_2(x)\neq 0$ exist
also away from the BPS limit and that in particular, the solutions
with $f_2(x)\neq 0$ are lower in energy than the corresponding embedded Abelian--Higgs strings.

\subsection{Globally regular semilocal strings}
We have solved the equations (\ref{fieldeq1}), (\ref{fieldeq2}) 
subject to the boundary conditions (\ref{bc}) numerically  using the ODE solver COLSYS \cite{colsys}.

The total energy and deficit angle don't change when varying the value of $f_2(0)$ for fixed
$\alpha$ and $n$. This is related to the zero mode present in the model
that persists to exist when studying the solutions in curved space-time.
A typical gravitating semilocal string solution is shown in Fig.~1 (left) for
$\alpha=0.5$ and $n=1$.


\begin{figure}
\includegraphics[width=8cm]{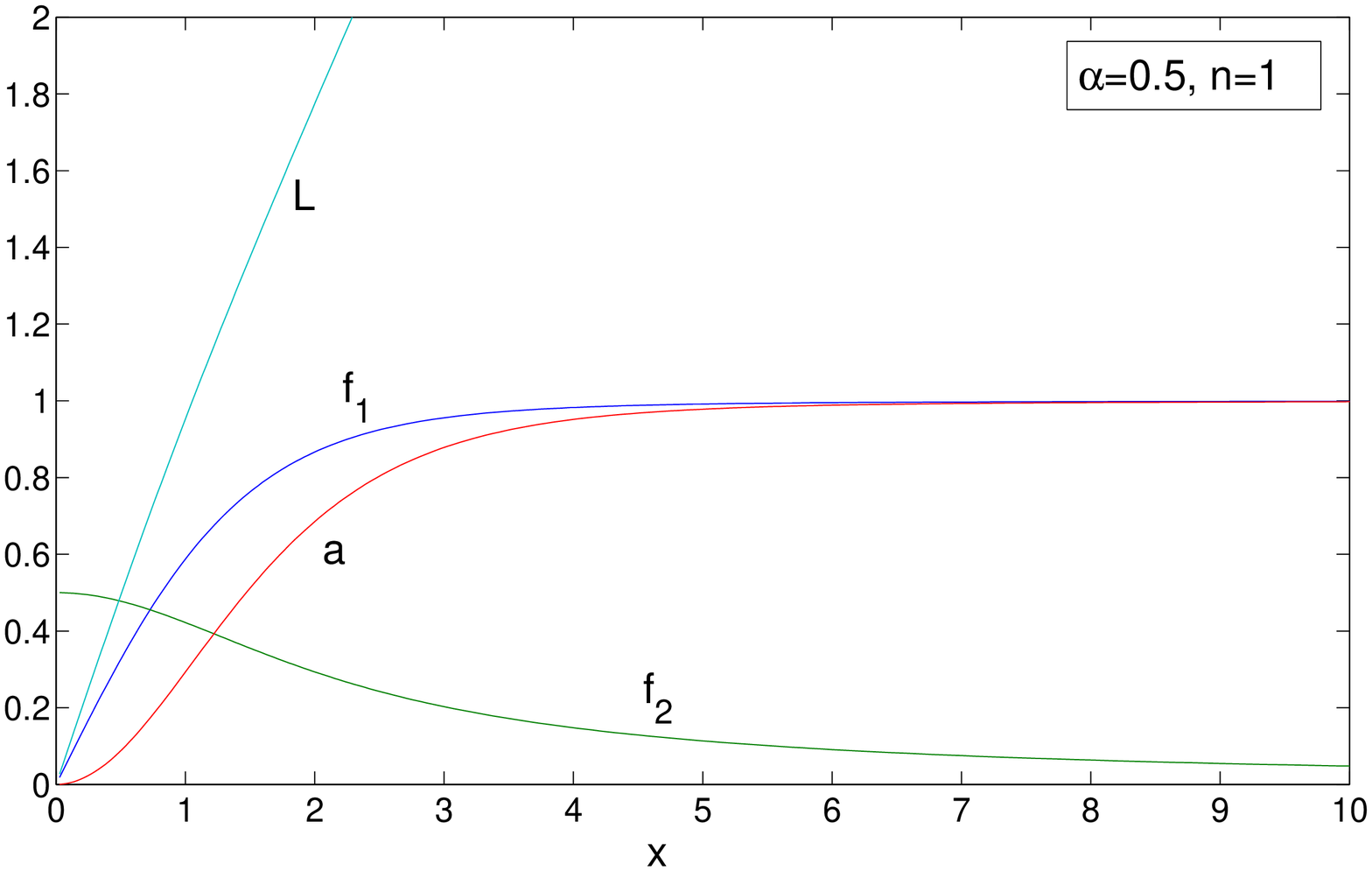} 
\includegraphics[width=8cm]{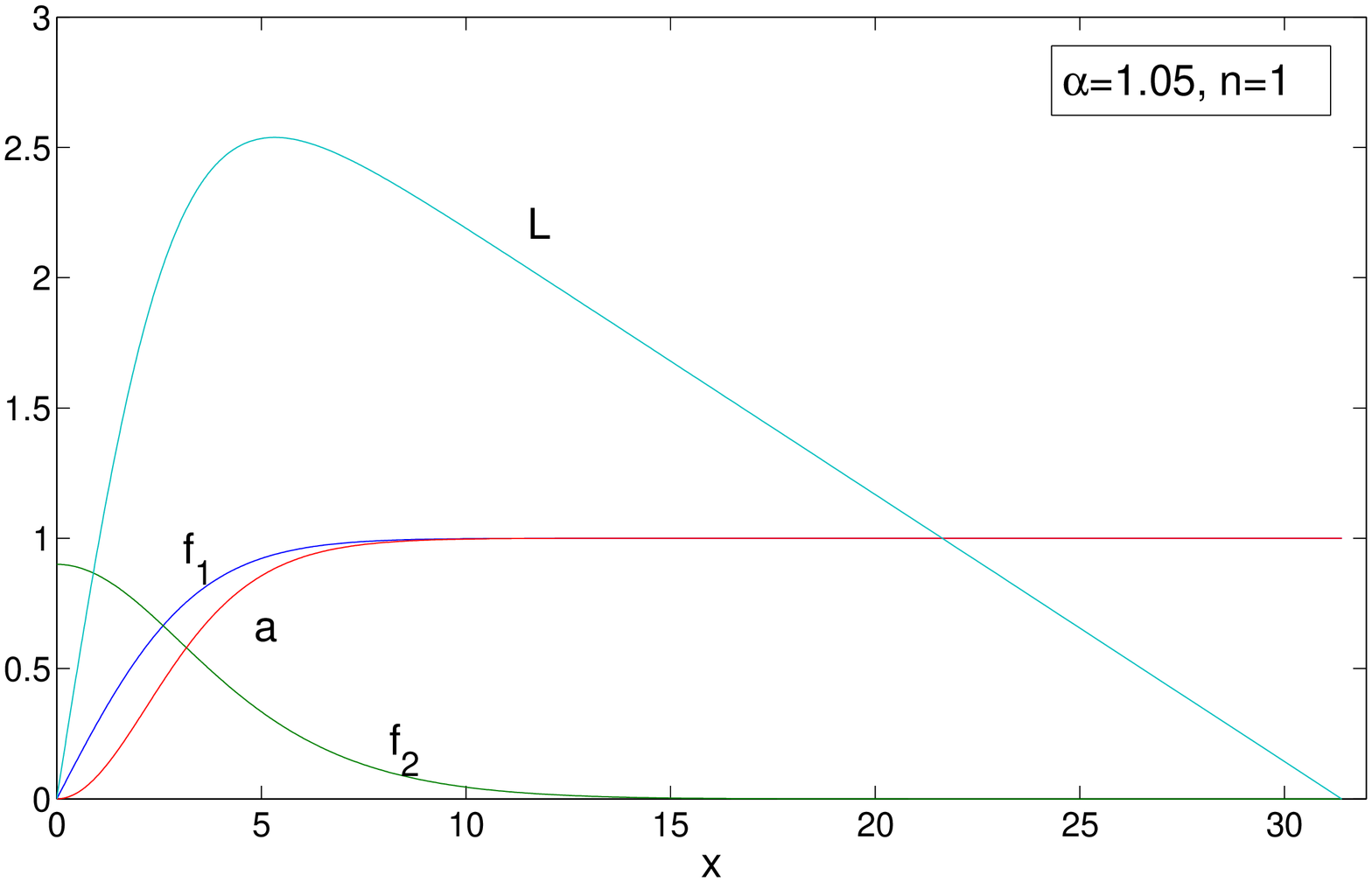}
\label{fig1}
\caption{The profiles of a typical gravitating semilocal string with $\alpha=0.5$ and $n=1$ (left)
and of a typical supermassive semilocal string with $\alpha=1.05$ and $n=1$ (right).}
\end{figure}

We have studied the dependence of the constant $c$ appearing in (\ref{relation}) on the value of
the condensate at the origin, $f_2(0)$.
For a fixed value of $f_2(0)$, we observe that $c$ decreases (slightly) for increasing $\alpha$
and that for $\alpha\rightarrow 1$, $c$ tends to a finite value.
Note that for $\alpha > 1$, the solutions become supermassive with $L(x_0)=0$ at a finite
value of the radial coordinate such that the integral in (\ref{relation}) diverges.
For a fixed value of $\alpha$, the value of $c$ depends strongly on the choice of $f_2(0)$.
For $f_2(0)\rightarrow 0$, the value of $c$ tends to infinity, while for $f_2(0)\rightarrow 1$ it
tends to zero.  This can be understood by looking at the non-gravitating ($n=1$) counterpart \cite{av}, where~:
\begin{equation}
\label{f1f2}
f_2(x)=\frac{1}{c}\frac{f_1(x)}{x}
\end{equation}

For $f_2(0)\rightarrow 0$ we have $f_2(x)\equiv 0$ and recover the embedded Abelian--Higgs solutions.
The relation (\ref{f1f2}) can only be fulfilled if $1/c\rightarrow 0$. For $f_2(0)\rightarrow 1$ we have $f_2(x)\equiv 1$, hence for the requirement of finiteness of the energy $f_1(x)\equiv 0$. From (\ref{f1f2}) we find that
$c\rightarrow 0$ in this limit.

\subsection{Supermassive semilocal strings}
For $\alpha^2 n > 1$, the metric function $L$ possesses a zero at a finite distance $x_0$
from the string core such that the solution
exists only on the interval $x\in [0:x_0]$. A typical solution is shown in Fig.~1 (right) for $\alpha=1.05$, $n=1$
and the value of the condensate $f_2(0)=0.9$. In this case $x_0\approx 31.4$.

\begin{figure}[htbp]
\centering
    \includegraphics[width=10cm]{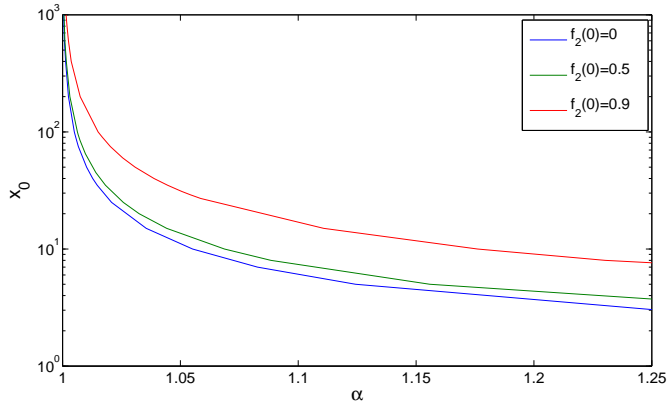} 
\caption{The value of the radial coordinate at which the metric function $L$ vanishes, $L(x_0)=0$
in dependence on the gravitational coupling $\alpha$ for $n=1$ and different
choices of the value of the condensate $f_2(0)$.}
\label{fig_sm2}
\end{figure}

We have studied the dependence of $x_0$ on $\alpha$ and the choice of the value
of the condensate on the string, $f_2(0)$. Our results are shown
in Fig.\ref{fig_sm2} for $n=1$. For comparison, we also give the values for
the embedded Abelian case with $f_2(0)=0$, i.e. $f_2(x)\equiv 0$.
For all choices of $f_2(0)$, $x_0$ tends to infinity for $\alpha\rightarrow 1$, indicating
that for $\alpha^2 < 1$, the solutions are globally regular.
For increasing $\alpha$, $x_0$ decreases. We observe that for a fixed value of $\alpha$
the value of $x_0$ increases with increasing $f_2(0)$, i.e.
the larger the condensate the larger the range of $x$ on which the supermassive
solutions exist.

\vspace{1cm}

In the limit $f_2(0)=1$, the solutions are such that $f_1(x < \infty)\equiv 0$, $f_2(x<\infty)\equiv 1$,
$a(x<\infty)\equiv 0$, while $f_1(x = \infty)= 1$, $f_2(x=\infty)= 0$,
$a(x=\infty)= 1$. This means that we have an infinite contribution to the energy density due to the infinite
derivatives of the matter functions at infinity. Hence, we need to require $L(x=\infty)=0$ in order to get
finite energy solutions. The increase of the condensate to $f_2(0)=1$ thus allows to make the
solutions regular on the interval $x\in [0:\infty[$ independent of the choice of the gravitational
coupling $\alpha$.

 \section{Conclusions}
We have studied the gravitational properties of semilocal strings in the BPS limit.
Like their non-gravitating counterparts, gravitating semilocal strings fulfill an energy bound
and a direct relation between the two components of the scalar doublet exists, which now depends
on the metric. The interval of the radial coordinate on which semilocal strings
with energy per unit length (roughly) larger than the Planck mass $\mu > M_{Pl}^2/4$  exist increases
with increasing value of the condensate and extends to infinity in the limit where the condensate tends
to unity.
\\
\\

\end{document}